\documentclass[sigplan]{acmart}
\pdfoutput=1

\usepackage{amsmath}
\usepackage{xspace}
\usepackage{comment}
\usepackage{graphicx}
\usepackage{url}
\usepackage{todonotes}
\usepackage{fancyvrb}
\usepackage{natbib}
\usepackage{titlesec}
\usepackage{setspace}
\titlespacing\section{0pt}{4pt plus 4pt minus 2pt}{0pt plus 2pt minus 2pt}
\titlespacing\subsection{0pt}{4pt plus 4pt minus 2pt}{0pt plus 2pt minus 2pt}

\newcommand{\sys}{{\tt Curator}\xspace}

\acmConference[TaPP'18]{USENIX Theory and Practice of Provenance}{July 9--13, 2018}{London, England}
\acmYear{2018}
\acmISBN{} % \acmISBN{978-x-xxxx-xxxx-x/YY/MM}
\acmDOI{} % \acmDOI{10.1145/nnnnnnn.nnnnnnn}
\startPage{1}

\title{\sys: Provenance Management for Modern Distributed Systems}

\author{Warren Smith}
\affiliation{%
\institution{The Weather Company, an IBM Business}
\streetaddress{400 Minuteman Rd}
\city{Andover}
\state{MA}
\postcode{01810}
\country{USA}}
\author{Thomas Moyer}
\affiliation{%
\institution{UNC Charlotte}
\department{Software and Information Systems}
\streetaddress{333G Woodward Hall\\9201 University City Blvd}
\city{Charlotte}
\state{NC}
\postcode{28223}
\country{USA}}
\author{Charles Munson}
\affiliation{%
\institution{MIT Lincoln Laboratory}
\department{Secure Resilient Systems and Technology}
\streetaddress{244 Wood St.}
\city{Lexington}
\state{MA}
\postcode{02421}
\country{USA}}

\titlenote{DISTRIBUTION STATEMENT A. Approved for public release: distribution unlimited.\\This material is based upon work supported by the Assistant Secretary of Defense for Research and Engineering under Air Force Contract No. FA8721-05-C-0002 and/or FA8702-15-D-0001. Any opinions, findings, conclusions or recommendations expressed in this material are those of the author(s) and do not necessarily reflect the views of the Assistant Secretary of Defense for Research and Engineering.}

\titlenote{ \noindent Permission to make digital or hard copies of all or part of this work for personal or classroom use is granted without fee provided that copies are not made or distributed for profit or commercial advantage and that copies bear this notice and the full citation on the first page. \\
\vspace{2pt}
\noindent \textsl{TaPP 2018}, \quad July 11--12, 2018, London, UK. \\
\noindent Copyright remains with the owner/author(s).}

\begin{document}

%!TEX root = curator-v2.tex

\begin{abstract}

Data provenance is a valuable tool for protecting and troubleshooting distributed systems. Careful design of the
provenance components reduces the impact on the design, implementation, and
operation of the distributed system.  In this paper, we present \sys, a provenance management
toolkit that can be easily integrated with microservice-based systems and other modern distributed
systems. This paper describes the design of \sys and discusses how we have used \sys to add provenance to
distributed systems. We find that our approach results in no changes to the design of these
distributed systems and minimal additional code and dependencies to manage. In addition, \sys uses the same
scalable infrastructure as the distributed system and can therefore scale with the
distributed system.

\end{abstract}

% 2012 ACM Computing Classification System (CSS) concepts
% Generate at 'http://dl.acm.org/ccs/ccs.cfm'.
\begin{CCSXML}
%<ccs2012>
%<concept>
%<concept_id>10011007.10011006.10011008</concept_id>
%<concept_desc>Software and its engineering~General programming languages</concept_desc>
%<concept_significance>500</concept_significance>
%</concept>
%<concept>
%<concept_id>10003752.10010124.10010138.10010143</concept_id>
%<concept_desc>Theory of computation~Program analysis</concept_desc>
%<concept_significance>300</concept_significance>
%</concept>
%</ccs2012>
\end{CCSXML}

%\ccsdesc[500]{Software and its engineering~General programming languages}
%\ccsdesc[300]{Theory of computation~Program analysis}
% end generated code

%\keywords
%keyword1, keyword2

%% Note: \maketitle command must come after title commands, author
%% commands, abstract environment, Computing Classification System
%% environment and commands, and keywords command.
\maketitle

%!TEX root = curator-v2.tex

\section{Introduction}
\label{sec-intro}

% area: data provenance is awesome!
Data provenance, the history of data as it moves through and between systems, provides distributed system
operators with a potentially rich source of information for a wide-range of uses.  Operators can use
provenance for troubleshooting~\cite{prov:zfn+2011}, auditing~\cite{prov:security:hsw2009}, and forensic
analysis~\cite{prov:lzx2013}.  Existing systems have focused on the collection and usage of provenance data, often
with the analysis occurring on the same system that collected the provenance.  While this works well for
applications running on a single host, it quickly breaks down on distributed systems.

% problem: distributed systems managmement of provenance
In systems that have considered provenance for distributed systems, the proposed architectures treat
provenance as unique from other sources of metadata, such as log and audit data.  This requires users
of provenance to build entire infrastructures to manage the provenance data, increasing the complexity of the
application and system.  This increased complexity can also increase the attack surface of the application,
negating the security benefits of adding provenance to a system.  What is needed is a provenance
management system that works in concert with existing infrastructures, and provides lightweight integration
into applications.

% solution: curator
In this paper, we present \sys, a provenance management toolkit that integrates with existing
logging/auditing systems.  Additionally, \sys provides a lightweight library to integrate provenance
collection into existing applications that minimizes dependencies, reducing the integration complexity
for application developers.  \sys is able to integrate provenance from multiple levels of abstraction,
including applications, infrastructure (databases, processing engines, etc.), and operating systems.  The
system ensures a consistent encoding of data between provenance sources, allowing consumers of provenance data
to reason about system behavior across different levels of the system.

We focus our attention on the integration of \sys into microservice-based systems, where applications
consist of small services that coordinate to achieve the goals of the application.  Such
architectures are popular in today's systems and present challenges when adding provenance.

%!TEX root = curator-v2.tex

\section{Design}
\label{sec-design}

The goals for the design of the \sys toolkit emerged from our experiences adding data provenance to prototype
data processing systems.

{\em G1) {\bf Minimally invasive:}} Our first goal is to make it easier to create and emit provenance from application
services and infrastructure. It was often difficult to add and de-conflict packages used by our previous
provenance instrumentation with the package dependencies of the data processing systems we were
instrumenting.

{\em G2) {\bf Scalable:}} The second goal is to aggregate and store provenance information in a scalable way while
re-using the infrastructure deployed by a data processing system. Maintainers of data processing systems are often reluctant to add additional software
infrastructure solely for the use of a data provenance subsystem, even when the system generates high volumes of provenance data.

{\em G3) {\bf Visualization:}} Our third goal is to provide a set of tools and displays for visualizing and analyzing provenance
information. We found that there are commonalities in the provenance visualization and analysis needs of the
systems we added data provenance to and that we could create components for use across these systems.

\begin{figure}
\centering
\includegraphics[width=0.40\textwidth]{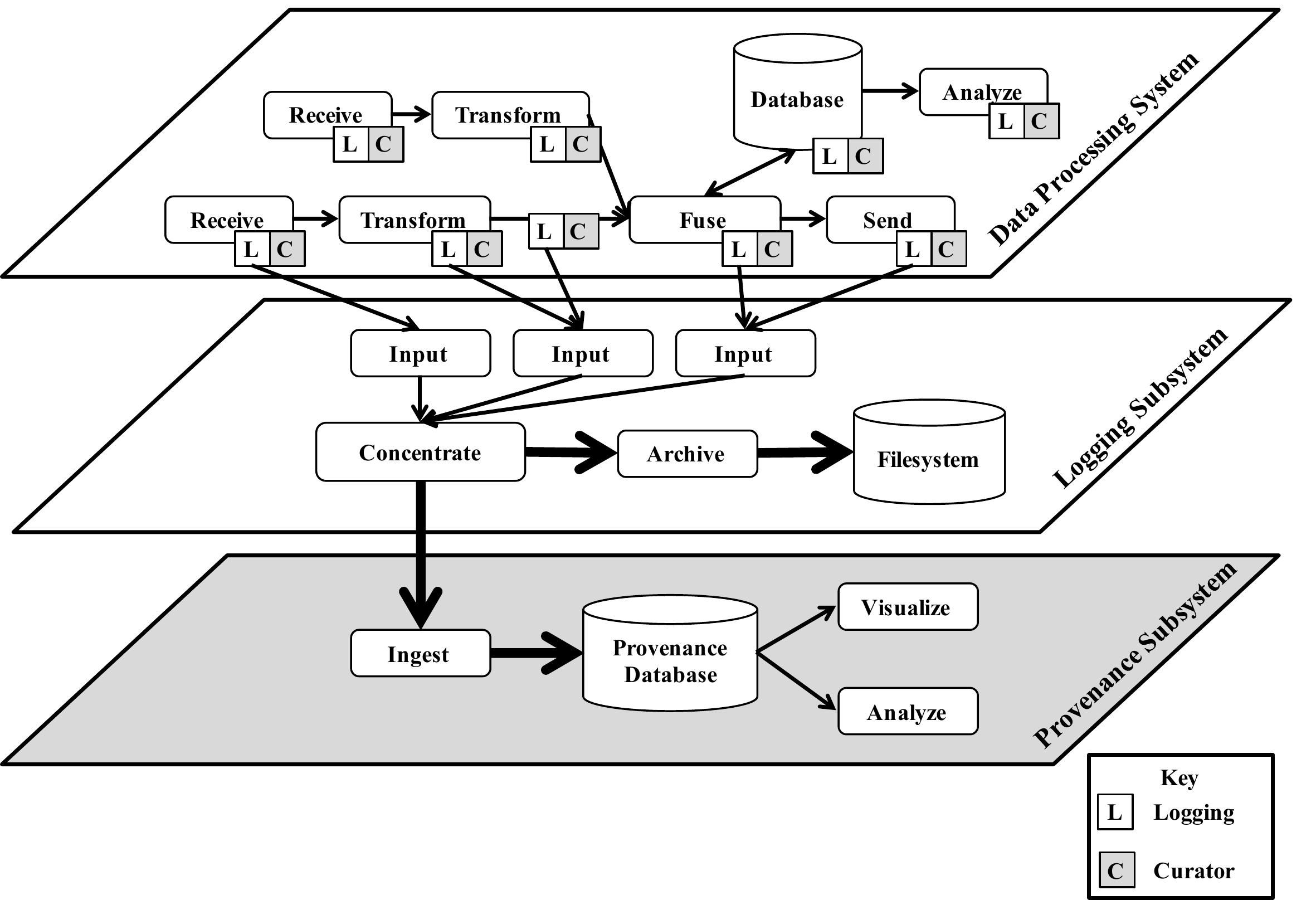}
\caption{The \sys toolkit integrated into a microservice architecture. Some connections have been removed for legibility.}
\label{fig-arch}
\end{figure}

Fig.~\ref{fig-arch} shows our approach to satisfy these goals. The key concept of this architecture is to
view gathering, storing, and analyzing provenance information as a logging problem. The \sys design consists
of helper libraries to create provenance information and add it to logs, use of a log aggregation and
processing system, and a provenance subsystem for storing, visualizing, and analyzing the provenance.

\subsection{Instrumentation and Logging}

To emit provenance, the \sys toolkit includes a Java library to create, encode, and log
provenance. Fig.~\ref{fig-emit} shows the creation of a provenance logger with a serializer, the creation of
provenance objects, and the logging of these objects to the same log the microservice uses for its
logging.

\sys adopts the World Wide Web Consortium's W3C-PROV~\cite{w3c-prov-overview} standards to represent data
provenance information and defines a set of Java objects organized as graph vertices and edges to represent
the PROV data model. The middle of Fig.~\ref{fig-emit} contains examples of these objects. To translate
provenance objects to and from representations that are suitable for logs, the toolkit contains
serializers and deserializers that support formats such as PROV-JSON~\cite{w3c-prov-json}. The third line of Fig.~\ref{fig-emit} demonstrates selection of a serializer.

% more description of W3C PROV? Fig. from the primer?

\sys makes use of the logging framework chosen by the authors of the microservice, rather than imposing a
logging framework, and currently supports log4j~\footnote{\url{https://logging.apache.org/log4j}} and log4j2. The toolkit also provides
wrapper {\tt ProvenanceLogger} classes to hold a serializer object and to implement {\tt log()} methods for all
of the provenance objects.

\begin{figure}
\begin{Verbatim}[fontsize=\footnotesize]
ProvenanceLogger logger =
  new ProvenanceLogger(Logger.getLogger("App"),
                     new ProvJsonSerializer());
Entity input = new Entity();
input.setAttribute("filename", "IMG-0942.jpg");
Activity transform = new Activity();
Used used = new Used(transform, inputData);
logger.log(inputData);
logger.log(transform);
logger.log(used);
\end{Verbatim}
\caption{Adding provenance instrumentation to a microservice.}
\label{fig-emit}
\end{figure}

\subsection{Aggregation and Ingest}

Once the microservices log provenance information, the logging system gathers the logs and then filters the
provenance information out of the logs. \sys deserializes and writes the provenance data into a provenance
database. \sys includes a simple service to perform these tasks. This service currently receives
log4j data via a socket or a log file. It then deserializes the provenance data using any of the
available deserializers (e.g PROV-JSON), and writes the provenance information to any supported provenance
database.

However, rather than using the simple \sys ingest service, we recommend using a log management system for a
more scalable and robust ingest implementation.  Systems such as logstash~\footnote{\url{https://www.elastic.co/products/logstash}} and
fluentd~\footnote{\url{https://www.fluentd.org/}} are widely used in microservice architectures to manage log data. These services are
modular and configurable and support the dual use of monitoring and analyzing the operation of a
microservice-based system as well as managing provenance data. The \sys toolkit currently includes a
logstash output plugin.

\subsection{Storage and Query}

We have found that it is not possible to select a single data storage solution for provenance data. First, the
volume and velocity of provenance data varies from system to system so no one solution is always the most
appropriate. Second, a microservice system has likely adopted a database for their needs and the developers
and operators would prefer to also use that database for provenance data.

The \sys approach is to therefore support a number of different databases behind common {\tt Store} and {\tt
  Query} interfaces. \sys currently supports popular SQL databases (MySQL/MariaDB, PostgreSQL, H2, Derby) and
the Accumulo~\footnote{\url{https://accumulo.apache.org}} distributed key/value store.  \sys represents provenance information as graphs of
vertices and edges that have attributes (key/value pairs). \sys stores these graphs in SQL in a normalized
form and in Accumulo in a denormalized form, as is typical for such databases. The optimized schemas used in
the databases enable fast retrieval of vertices and edges by their ids and locating vertices and edges that
have specific attributes. It has been demonstrated that SQL databases support lower ingest rates and data
volumes, but fast queries\footnote{This assumes that the tables have the appropriate indexes.}. Accumulo
supports high ingest rates and volumes of data, but queries can be slower~\cite{moyer2016high}.

The \sys~{\tt Query} interface supports a number of operations to retrieve provenance data to drive
analytics. As mentioned above, this interface supports basic operations such as finding vertices and
edges by identifier or by attributes. The interface also supports finding ancestors and descendants of a
vertex (typically an entity) so that an analysis tool can determine what entities, activities, and agents
influenced or were influenced by a vertex. For broader views, the {\tt Query} interface supports finding the
ancestors and descendants of a set of vertices and the connected subgraph that a set of vertices are part of.

%\subsection{Analytics and Response}
\subsection{Analytics}

We use the provenance information available from the {\tt Query} to drive analytics. For example, we have
written system-specific analysis code to analyze the structure and content of a provenance graph related to a
data item passing through a data processing pipeline in order to ensure accurate and timely processing of data inputs. This code determines if the provenance graph has the
structure and content that we expect and notifies the user if it does not. Such analysis can be used for anomoaly detection and for debugging. However, in the case of anomaly detection, more work is needed to ensure the security of the provenance collected using Curator.
Graph-grammar based techniques such as those used in Winnower can be integreated to allow for more general-purpose analysis of graph structure~\cite{hbm2018}.

\subsection{Visualization}

Finally, provenance visualizations are useful for a number of purposes. An operator or developer can use
visualizations to understand the operation of a system and the interactions of the components in the
system. They can also use visualizations to analyze the causes or effects of a failure or an attack.
\sys includes visualization components for integration into data processing systems.
These components provide a general-purpose mechanism to explore provenance data that has been processed by \sys, allowing consumers of provenance data to explore the provenance data in a browser.

%!TEX root = curator-v2.tex

\section{Use Cases}
\label{sec-use-cases}

We used the \sys toolkit to add provenance to several distributed
systems. The general purpose of all of these systems is to process data as it streams into the system, store
data products, perform analyses on stored data, and present results to users. The systems are
architected as a set of microservices that communicate via a message bus.

We integrated \sys into these systems in order to collect provenance. First, we added
provenance logging statements into the application services. As described above, these statements are not
difficult to write and the application services have a great deal of contextual information that can be of use
in the provenance record. However, it is challenging to locate all of the code locations that require logging. One promising technique for auotmatically instrumenting applications for provenance is described in~\cite{capobianco_tapp17}. Such systems can use \sys to add provenance to legacy applications.

Second, we added provenance logging to common application libraries, such as a library that wraps a messaging
system to package application data for sending and receiving. This is again easy to accomplish using code
similar to that shown in Fig.~\ref{fig-emit}. This approach works well because provenance logging is added in a minimal number of locations and can gather a large amount of the provenance needed for a system. In addition,
because this logging is part of the application, a large amount of application context is available to the provenance system.

Third, we added provenance logging to common infrastructure, such as Spring Integration. Spring
Integration~\footnote{\url{https://projects.spring.io/spring-integration}} is a Java framework for creating distributed applications by composing
services and abstracting the message-oriented mechanisms that connect the services. This approach allows
services to be written independently from each other focusing only on their inputs, functionality, and
outputs. To add provenance gathering to Spring, we created a Spring Interceptor that creates provenance log
entries for each message it sees. We also made a small change to the Spring XML configuration to add
this interceptor to every channel so that we could gather provenance information for every communication
between services. More details on the Spring Interceptor can be found in the full technical report. Details of the technical report can be found at \url{https://bitbucket.org/crestlab/Curator-public.git}.

%!TEX root = curator-v2.tex

\section{Related Work}
\label{sec-rel-work}

% should have some text above if we want to start a subsection here
%\subsection*{Provenance Management Systems}
\sys is not the first provenance management toolkit. SPADE~\cite{prov:gt2012} and PLUS~\cite{prov:cab+2011} are two systems that provide similar capabilities. These management systems aim to capture, encode, store, and query provenance information from a wide-range of systems. PLUS aims to support integration into applications by providing a library for reporting provenance, a SQL schema for storing provenance, and a query interface to access collected provenance. PLUS also provides a modified Mule enterprise service bus that captures provenance and a provenance reporting API.
% a quick google seems to indicate that it is 'Mule', not 'MULE'
% add a citation for it?

SPADE is a similar system that aims to collect, encode, store, and query provenance data in distributed systems. The core of SPADE is a pluggable kernel with interfaces for collection, storage, and querying of provenance data. SPADE also considers how to handle provenance at multiple abstraction layers, with collection plugins to capture both operating system information from audit logs and from applications using a named pipe. Similar to PLUS, SPADE provides an API for applications to report provenance. In both SPADE and PLUS, the transfer of provenance data is handled separately from other logging messages.

The primary difference between \sys and the tools described above is that \sys integrates with existing logging infrastructures for distributed systems, instead of building an entirely parallel infrastructure to support the management of the provenance data. This provides the additional benefit that system designers and developers can integrate with their existing logging infrastructures. This also reduces the complexity of \sys, since it is not responsible for log management. Our initial version of \sys acted as a centralized log management system for provenance data, making the core of \sys large and complex.

%\subsection*{Provenance Libraries}
ProvToolbox~\footnote{\url{http://lucmoreau.github.io/ProvToolbox}} is a Java library for working with provenance in the W3C-PROV format. It defines
Java classes to represent provenance documents. ProvToolbox also provides functionality
to convert between Java provenance documents and the W3C PROV formats (PROV-XML, PROV-JSON, PROV-N). Finally,
it can generate images of provenance graphs using Graphviz~\footnote{\url{http://www.graphviz.org}}.
However, we found that ProvToolbox didn't
fulfill all of our needs. First, ProvToolbox depends on a large number of packages making it
challenging to integrate with microservices that have their own, sometimes conflicting
dependencies. Second, we desired an in-memory graph representation of provenance information that we could
easily search and traverse and that would match the representation that we store in databases, which ProvToolbox does not provide. Third, we wanted a simpler
API based on the expectation that we will need to implement that API
in a variety of programming languages. These reasons led us to develop the \sys representation for provenance
graphs and support for serializing and deserializing those graphs to W3C-PROV formats.

The Provenance-Aware Storage System, PASS~\cite{prov:mbh+2009} supports an API for disclosing provenance called the Disclosed Provenance API, or DPAPI. The Core Provenance Library, CPL~\cite{prov:ms2012} also provides an API for reporting provenance from userspace applications. The PASS kernel and userspace applications utilize the DPAPI to integrate provenance from the OS and from the applications into a single provenance record for the system. Unlike ProvToolbox, the DPAPI and CPL do not require a particular specification. Instead, they are adaptable to different specifications. However, the DPAPI also expects the system to be running the PASS kernel, which may not be feasible for all applications.

%!TEX root = curator-v2.tex

\section{Future Work}
\label{sec-future}

Several open challenges still exist that Curator does not currently address. First, security of collected provenanve data. Sevreral systems have looked at ways to provide secure provenance~\cite{prov:security:hsw2009,btb+2015} and it remains to be seen how Curator integrates with these. Second, linking high-level, semantically rich provenance to low-level system provenance continues to be a challenge for all provenance systems that do not provide explicit linking between high- and low-level events~\cite{prov:mbh+2009}. As we continue to leverage Curator, these enhanced capabilities will be explored further.

Our research and development roadmap also includes adding a number of analytics and visualization capabilities to \sys.
We plan to add support for streaming analytics so that provenance data is analyzed as it arrives.
In both of these areas, we will follow our existing approach and use existing systems to provide the primary functionality (e.g. Storm~\footnote{\url{https://storm.apache.org}}) with \sys providing provenance-specific components to use with those systems.
Future work towards visualization of provenance data includes an interactive provenance graph, allowing users to navigate through provenance data visually, as well as to filter and search the provenance data to find nodes or portions of the graph that match given criteria.

%!TEX root = curator-v2.tex

\section{Conclusion}
\label{sec-conclude}

This paper described \sys, a toolkit designed to  make it easier to integrate provenance into distributed
systems. \sys provides tools for logging provenance, retrieving provenance from logs, storing provenance in
databases, and analyzing and visualizing provenance. \sys is a composable set of tools where a developer
can select the tools, such as provenance loggers, that best match their system. This paper also provided examples of how we used \sys in distributed systems. We have found that the design of
\sys increases the likelihood that developers will add provenance to a system because it lowers the impact on
that system compared to previous approaches. We also found that \sys reduces the effort it takes to add
provenance to a system. Existing systems for generating and managing provenance information do not have these
advantages.

\section*{Availability}
The source code for Curator can be found at \url{https://bitbucket.org/crestlab/Curator-public.git}.

\begin{acks}                            %% acks environment is optional
                                        %% contents suppressed with 'anonymous'
The authors would like to thank the anonymous reviewers for their helpful feedback. They would also like to acknowlege the hard work of the Secure Resilient Systems and Technology Group at MIT Lincoln Laboratory, many of whom had discussions with the authors about what would make Curator most useful for their work.
\end{acks}

{\footnotesize
\bibliographystyle{acm}
\bibliography{citations}
}

%!TEX root = curator-v2.tex
\newpage
\onecolumn
\appendix
\section{Spring Integration Channel Interceptor}
\label{app:interceptor}
Fig.~\ref{fig-interceptor} shows a Spring Interceptor for provenance that allows us to create {\tt
  wasDerivedFrom} relationships between messages as the messages flow through the system. In addition, we use
thread local storage to share information between the interceptor and the microservices. The interceptor saves
the message identifier of the message being handled and a microservice can then create {\tt used} or {\tt
  wasDerivedFrom} relationships to {\tt entity}s contained within the message. The microservice also records
the {\tt entity}s accessed while handling a message and then coordinates with the interceptor to add these
{\tt wasDerivedFrom} relationships for the outgoing message to the provenance log.

\begin{figure}
{\footnotesize
\begin{verbatim}
public class ProvChanInter extends ChannelInterceptorAdapter {
  @Override
  public Message<?> preSend(Message<?> message,
                            MessageChannel channel) {
      Entity msg = new Entity(message.getHeaders().getId());
      for (String name : message.getHeaders()) {
           msg.setAttribute(name, message.getHeaders()
                                         .get(name)
                                         .toString());
      }
      logger.log(msg)
      if (message.getHeaders().containsKey("previousId")) {
          logger.log(new WasDerivedFrom(message.getHeaders()
                                        .getId(),
                                          message.getHeaders()
                                        .get("previousId")));
      }
      return MessageBuilder.fromMessage(message)
                           .setHeader("previousId",
                              message.getHeaders().getId()
                           .toString()).build();
  }
}
\end{verbatim}
}
\caption{Spring channel interceptor to log provenance.}
\label{fig-interceptor}
\end{figure}

\newpage
\section{Graphical Interface for Visualization Tool}
\label{app:ui}
\begin{figure}
\includegraphics{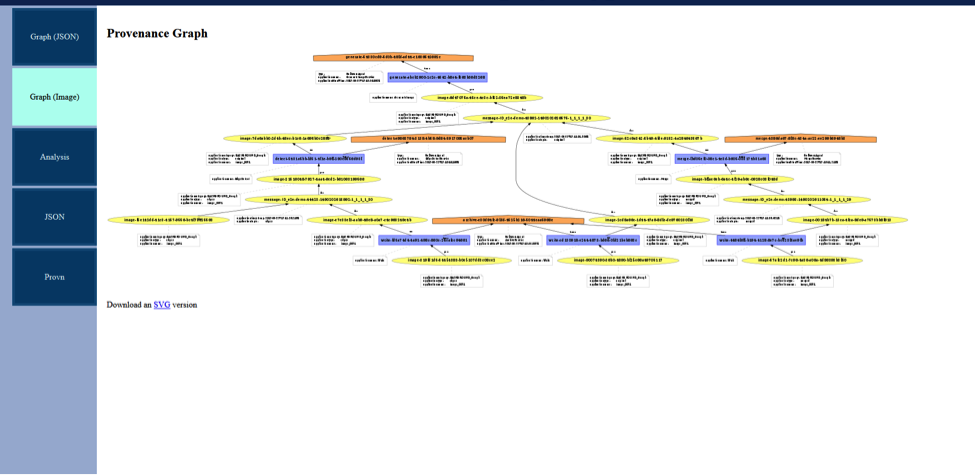}
\caption{Example of the visualizer web application.}
\label{fig:visualizer}
\end{figure}

Figure~\ref{fig:visualizer} shows the web application that can display provenance graphs.

\end{document}